\documentclass[aps,prl, showpacs, reprint,superscriptaddress]{revtex4-1}

\usepackage{graphicx}
\usepackage{amssymb,amsmath}
\usepackage{bm}% bold maths
\usepackage{float}
\usepackage{xcolor}
\definecolor{midnightblue}{cmyk}{1,1,0,0.1}
\definecolor{forestgreen}{cmyk}{0.75,0,1,0.5}

\usepackage{hyperref}
\hypersetup{
    bookmarks=true,         % show bookmarks bar?
    unicode=false,          % non-Latin characters in AcrobatÕs bookmarks
    pdftoolbar=true,        % show AcrobatÕs toolbar?
    pdfmenubar=true,        % show AcrobatÕs menu?
    pdffitwindow=false,     % window fit to page when opened
    pdfstartview={FitH},    % fits the width of the page to the window
    pdftitle={My title},    % title
    pdfauthor={Author},     % author
    pdfsubject={Subject},   % subject of the document
    pdfcreator={Creator},   % creator of the document
    pdfproducer={Producer}, % producer of the document
    pdfkeywords={keyword1} {key2} {key3}, % list of keywords
    pdfnewwindow=true,      % links in new window
    colorlinks=true,       % false: boxed links; true: colored links
    linkcolor=midnightblue,          % color of internal links (change box color with linkbordercolor)
    citecolor=magenta,        % color of links to bibliography
    filecolor=midnightblue,      % color of file links
    urlcolor=midnightblue,          % color of external links
}

\begin{document}

\title{Designer Topological Insulators in Superlattices}

\author{Xiao Li}
\affiliation{International Center for Quantum Materials, School of Physics, Peking University, Beijing 100871, China}
\author{Fan Zhang}
\affiliation{Department of Physics and Astronomy, University of Pennsylvania, Philadelphia, PA 19104, USA}

\author{Qian Niu}
\affiliation{International Center for Quantum Materials, School of Physics, Peking University, Beijing 100871, China}
\affiliation{Department of Physics, University of Texas at Austin, Austin, TX 78712, USA}

\author{Ji Feng}
\email{jfeng11@pku.edu.cn}
\affiliation{International Center for Quantum Materials, School of Physics, Peking University, Beijing 100871, China}

\date{\today}

\begin{abstract}
Gapless Dirac surface states are protected at the interface of topological and normal band insulators.
In a binary superlattice bearing such interfaces,
we establish that valley-dependent dimerization of symmetry-unrelated Dirac surface states
can be exploited to induce topological quantum phase transitions.
This mechanism leads to a rich phase diagram that allows us to design strong, weak, and crystalline topological insulators.
Our \emph{ab initio} simulations further demonstrate this mechanism in $[111]$ and $[110]$ superlattices of calcium and tin tellurides.
\end{abstract}

%\keywords{}

\maketitle

\textcolor{forestgreen}{\emph{\textsf{Introduction}.}}---
A topological insulator (TI) \citep{HasanKane10, QiZhang10, Moore10} is an insulating material whose electronic Hamiltonian cannot be adiabatically deformed into that of an atomic insulator.
Consequently, a TI has a protected gapless Dirac state on its interface with a normal insulator (e.g., vacuum).
Owing to the bulk energy gap, the charge transport of a TI is dominated by the gapless surface states and is robust against non-magnetic disorder.
The coupling of TI surface state with a second material is especially intriguing.
Proximity coupling the TI surface state to an $s$ wave superconductor breaks $U(1)$ gauge symmetry to yield a Majorana excitation~\citep{Fu08, Zhang13b}.
A magnetic proximity effect can break time-reversal symmetry producing a chiral edge state~\citep{Yu10, Chang13, Zhang13c}.
Both effects offer promising routes to novel devices, such as energy-efficient electronics and topological quantum computing.

Tremendous effort has been invested into the search for novel topological phases~\citep{HasanKane10, QiZhang10, Moore10, Fu11, Fu12}.
It is highly desirable to design and synthesize these phases starting with  a few accessible materials or elements.
Artificially created superlattices provide a natural arena for such an endeavor,
which is well provided with chemical, structural, orbital, and spin degrees of freedom.
One aim of this paper is to establish that in a binary superlattice, independent dimerization of symmetry-unrelated interfacial states
can be exploited to induce topological quantum phase transitions without breaking any symmetry.
The proposed mechanism allows us to design various TIs by tuning valley-dependent interlayer couplings.
A second objective of this paper is to demonstrate, in combination with density-functional theory calculations,
that with this strategy a binary superlattice  of calcium and tin tellurides 
can be fashioned into strong, weak, and crystalline topological insulators (sTI, wTI and cTI, respectively)~\citep{Fu06, Fu07, Fu11, Fu12}.

\begin{figure}[b!]
\includegraphics[width=7.5 cm]{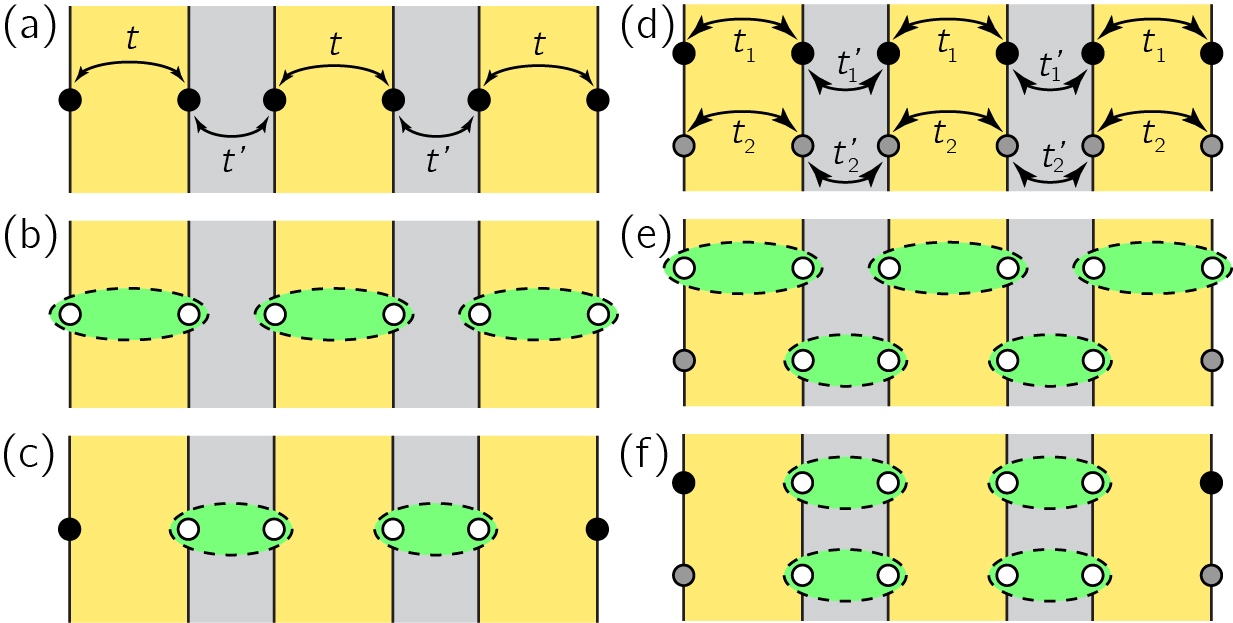}
\caption{Schematic depiction of surface-state dimerization.
(a) A superlattice composed of the parent sTI layers (yellow) and the spacer normal insulator layers (light grey).
(b) Dimerization through the parent layers. (c) Dimerization through the spacer layers.
(d) A superlattice with two independent Dirac interfacial states.
(e) and (f) show two different schemes of dimerization.
The black and gray circles represent isolated Dirac interfacial states
while the white states are gapped because of dimerization, as indicated by the green blobs.}
\label{fig:dimer}
\end{figure}

\textcolor{forestgreen}{\emph{\textsf{Surface-state dimerization}.}}---
We begin with an intuitive picture of the surface state polymerization, analogous to the Su-Shrieffer-Heeger model of polyacetylene~\citep{Su79}. Consider a superlattice formed by alternating layers of a parent sTI and a spacer normal insulator (NI), as shown in Fig.~\ref{fig:dimer}(a).
When the widths of parent and spacer layers are finite, there are two kinds of interlayer hoppings, $t$ and $t'$,
between the Dirac interfacial states. They account for, respectively, the covalent interaction between the pair of interfacial states through the parent and spacer layers. There appears to be a delicate interplay between $t$ and $t'$, as shown in Fig.~\ref{fig:dimer}. In the limit of $t \gg t'$, the interfacial states dimerize through the parent TI layers, annihilating the Dirac nodes at finite truncation of the superlattice. In the opposite limit of $t \ll t'$, interfacial states dimerize through NI layers, and it is inevitable that a Dirac surface state is still present on an outermost surface, as long as time-reversal symmetry is unbroken. Consequently, $t \ll t'$ leaves the superstructure a three-dimensional sTI.

Now we consolidate the above dimerization picture in an effective model. A Dirac interfacial state can be described by a Hamiltonian, $H_{v}({\bm k}) =  c_v {\bm q}_v \cdot \sigma$, where ${\bm q}_v = ({\bm k} - {\bm k}_v)_{||}$ is the crystal momentum component parallel to the cleaved surface, $c_v = \partial \varepsilon_v / \partial q$ is the Fermi velocity, and $\bm\sigma$ are the Pauli matrices of spin or pseudospin. We include an index $v$ in $H_{v}({\bm k})$, in anticipation of multiple Dirac valleys.
The surface-state dimerization is readily understood by the following Hamiltonian~\citep{Burkov11}
\begin{equation}
\mathcal{H}_v ({\bm k}) = E_{v}
+ \tau_z H_v
+ \tau_x (t_v + t'_v \cos k_z)
+ \tau_y t'_v \sin k_z,
\label{H}
\end{equation}
where $\tau_z=\pm$ denotes the two interfaces of a TI layer and $E_v$ reflects the possible particle-hole asymmetry of the TI.
We assume that the superlattice possess an inversion symmetry for simplicity.
The band dispersion of Hamiltonian~(\ref{H}) reads
$\varepsilon = E_v \pm (c_v^2q_v^2 + t_v^2+t_v'^2+2 t_v t'_v \cos k_z)^{1/2}$,
with a band gap $2||t_v|-|t'_v||$. We notice that this band gap closes at $k_z = 0$ ($\pi$) for $t_v' = -(+)t_v$,
which indicates topological quantum phase transitions in the presence of inversion symmetry.
The topological classification of the resultant phase depends on the relative strengths of $|t_v|$ and $|t'_v|$,
leading to two possible surface-state dimerization schemes, in agreement with the limiting cases shown in Fig.~\ref{fig:dimer}(b) and (c).  Therefore, tuning $|t'_v/t_v|$ represents a continuous pathway to convert a TI into a NI~\citep{Murakami07, Burkov11}.

When multiple Dirac interfacial states are present, the manifestation of interlayer coupling becomes more profound.
An event of band inversion corresponds to a topological quantum phase transition, accompanied by the gap closure.
The surface states of the parent TIs are located at the time-reversal-invariant points
in the ${\bm k}_{\parallel}$-space, and the Hamiltonian~(\ref{H}) has inversion ($\hat{\mathcal{P}}=\tau_x$) symmetry.
It follows that for a time-reversal-invariant point of ${\bm k}_{\parallel}$
\begin{equation}
\delta_v=\xi_v(0)\xi_v(\pi)=(-1)^{{\Theta}(|t'_v|-|t_v|)}
\label{eq:parity_exchange}
\end{equation}
determines the product of parity eigenvalues of all the occupied bands at $k_z = 0$ and $\pi$.
where $\Theta(x)$ is the Heaviside step function.
When different surface states are not related by any symmetry, they may dimerize in distinct manners, as depicted in Fig.~\ref{fig:dimer}(d).
We can further conclude that the overall $\mathcal{Z}_2$ invariant is~\citep{Fu07}
\begin{equation}
(-1)^{\nu_0} = \prod_v \delta_v.
\label{eq:nu0}
\end{equation}
Based on the criteria in Eqs.~(\ref{eq:parity_exchange}) and (\ref{eq:nu0}), the opportunity to design various TI phases is immediately evident. If $|t_1|>|t_1'|$ and $|t_2|<|t_2'|$, the superlattice is a sTI, as depicted in Fig.~\ref{fig:dimer}(e). If $|t_v'/t_v| > 1$ for both interfacial states, a pair of Dirac nodes still persist on each outermost surface, as shown in Fig. \ref{fig:dimer}(f). In the latter case the superlattice is either a wTI or a cTI.

\begin{figure}[t!]
\includegraphics[width=7.5cm]{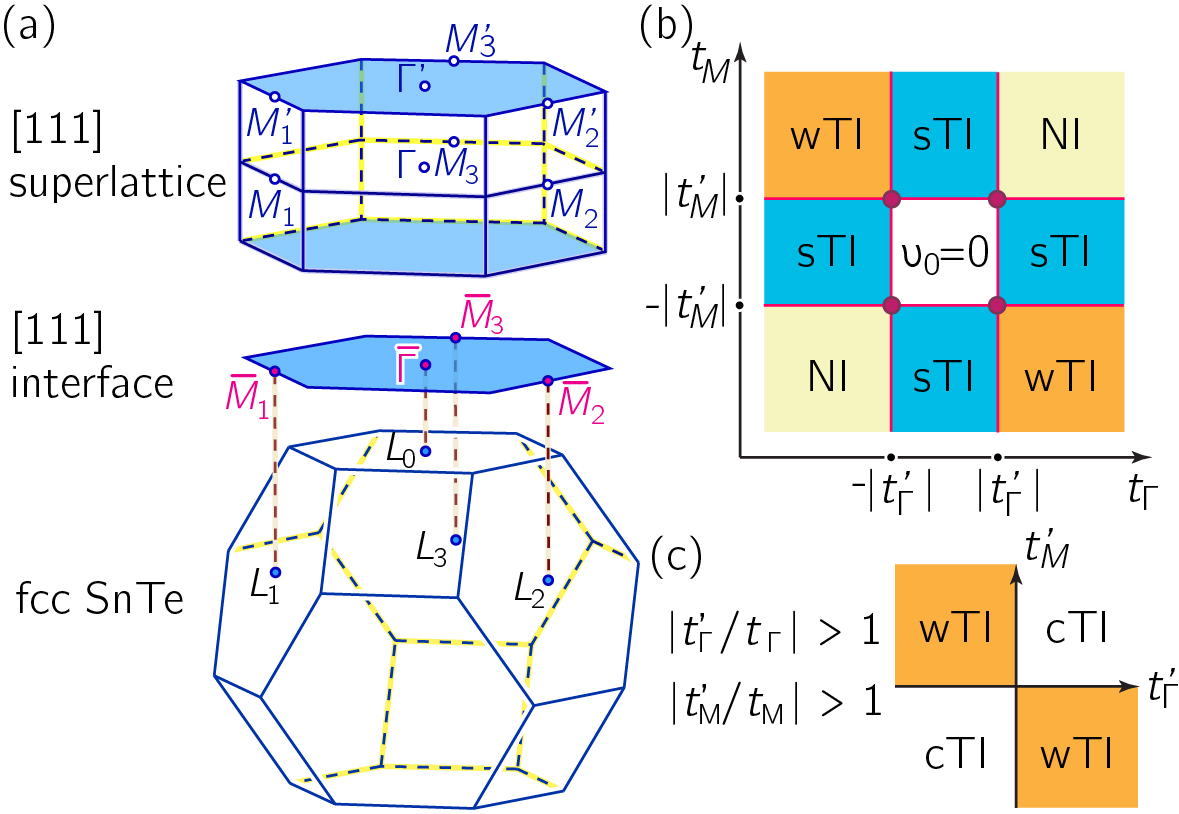} \caption{(a) Brillouin zone projection in [111] direction.  (b) Topological phase diagram of the superlattice. For a [111] interface of SnTe with a NI, the Dirac nodes are located at $\bar{\Gamma}$ and $\bar{M}_j \; (j=1,2,3)$ on the surface Brillouin zone. Extending to the superlattice, each valley branches out into a doublet at $k_z = 0$ and $\pi$, namely, $\Gamma, \Gamma'$ and $M_j, M'_j$. (c) The $v_0$ square in the center of (b).}
\label{fig:111}
\end{figure}

\textcolor{forestgreen}{\emph{\textsf{Superlattices of cTI}.}}---The above analysis suggests a unifying pathway to tailor-make three different classes of TIs through \emph{valley-dependent} interfacial state dimerization. In what follows, we present a concrete demonstration using the unique interfacial states of tin telluride (SnTe), a representative topological crystalline insulator. SnTe has the sodium chloride structure and $(110)$-like mirror symmetries. Band inversions occur near the four inequivalent $L$ points, as shown in Fig.~\ref{fig:111}(a), giving rise to multiple non-zero mirror Chern numbers~\citep{Fu12}. Consequently, an even number of Dirac surface states appear on any surface~\citep{Xu12, Tanaka12, Dziawa12} or interface that preserves a $(110)$-like mirror symmetry. We will focus on the $[111]$ and $[110]$ superlattices that harbor symmetry-distinct valleys.

The $[111]$ superlattice of SnTe with a normal insulator is particularly interesting, because in this case all the time-reversal invariant points of the superlattice are derived from the interfacial Dirac valleys, as explained in Fig. \ref{fig:111} (a). The parity analysis at three $M_j$ points are identical, and likewise at three $M'_j$ points, as lone as the $\hat{C}_3$ symmetry along the $[111]$ direction is unbroken. The topological class of the $[111]$ superlattice of SnTe follows readily from our arguments below Eq.~(\ref{eq:nu0}). The phase diagram is plotted in Fig.~\ref{fig:111}(b) and (c). As we can see in the $t_\Gamma$-$t_M$ parameter space, three topological phases --- the strong, the weak, and the crystalline --- can all be created in such superlattices.

As anticipated, when $\delta_{\Gamma}\delta_{M} = -1$, the resulting superlattice is a sTI (blue regions in Fig. \ref{fig:111}).
When $|t'/t|<1$ or $|t'/t|>1$  for both $\Gamma$ and $M_j$, the superlattice has an overall $\nu_0=0$.  Specifically, the classification of the band topology depends on the relative signs of  $t_\Gamma$ and $t_M$ for $|t'/t|<1$ in both valleys, where the superlattice can be either wTI with $\nu_0;(\nu_1\nu_2\nu_3)=0;(001)$ ($t_\Gamma t_M<0$) or a trivial insulator ($t_\Gamma t_M>0$). When $|t'/t|>1$ in both valleys, the classification depends on the signs of  $t'_\Gamma$ and $t'_M$, where the superlattice can be either wTI ($t'_\Gamma t'_M<0$) or congener cTI ($t'_\Gamma t'_M>0$) (Fig.  \ref{fig:111} (c)). It is worth remarking that the points where $|t'/t|=1$ for both $\Gamma$ and $M_j$, as indicated by the red circles in Fig. \ref{fig:111} (b), are \emph{topological tetracritical points} \citep{LiuFisher73, Fisher74}, which reflect the simultaneous presence of topological order of two distinct valleys. It may also be deduced that a finite neighborhood of the phase boundaries, including the tetracritical points, in Fig. \ref{fig:111}(b) is metallic via electron-hole compensation between valleys.

\begin{figure}[ht]
\includegraphics[width=8 cm]{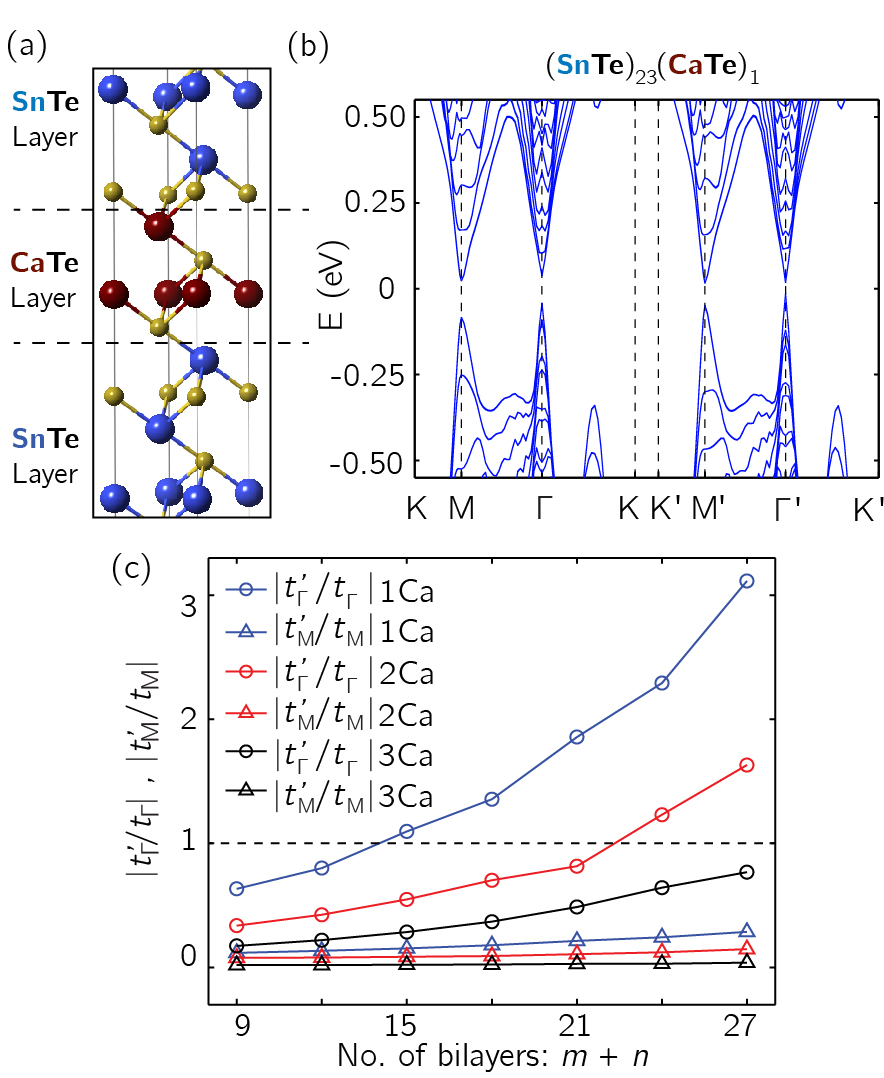}\caption{(a) A representative structure of $(m,2)_{[111]}$ superlattice. Only a section of the superlattice near the interface is shown. The blue, red and yellow balls correspond to Sn, Ca and Te, respectively. (b) The band structure near the band gap of $(23,1)_{[111]}$ superlattice. The Fermi level is set to zero energy. (c) The values of $|t'/t|$  as a function of the thickness of superlattice, at $\Gamma$ and $M$. Blue, red and black curves correspond to $n=1$, 2 and 3, respectively.}
\label{fig:band111}
\end{figure}

The valley engineering proposed above can only be established with judicious materials design. We employ density-functional theory simulations to study the superlattices of SnTe \citep{Perdew98, Kresse99}. Computational details can be found in Supporting Information (SI) \citep{SI}. We choose the isostructural calcium telluride (CaTe) for the spacer layer, instead of the obvious choice of isovalent IV-VI semiconductors, PbTe and GeTe, for two reasons. First, the lattice mismatch between SnTe and CaTe is less than 1\%, a feature conducive to experimental growth of heterostructures. Second, GeTe is known to undergo ferroelectric distortion \citep{Pawley66}, which, albeit interesting in itself in the context valley engineering, unduly complicates a first analysis. CaTe is a normal insulator with a computed gap of 1.3 eV. Our calculations show that SnTe has 107 meV direct gaps at $L$, in agreement with previous work \citep{Fu12}. A [111] superlattices composed of $m$ SnTe bilayers and $n$ CaTe bilayers is denoted $(m,n)_{[111]}$ for brevity. A representative structure of $(m,2)_{[111]}$ superlattice is shown in Fig. \ref{fig:band111} (a). Inversion symmetry is present in all superlattices considered, after structural optimization.

Our calculations examine $(m,n)_{[111]}$ with a single, double and triple bilayer of the normal insulator CaTe ($n=1,2,3$), with total number of bilayers up to $m+n=27$. A representative band structure for $(23,1)_{[111]}$ superlattice is shown in Fig. \ref{fig:band111} (b), which has an overall direct gap 30 meV. Based on the direct gaps of the Kohn-Sham states at the time-reversal invariant points in relation to Hamiltonian (1), we estimate the $|t|$ and $|t'|$ at $\Gamma$ and $M$. For a given valley, we take $4|t| = |\Delta(0)+\Delta(\pi)|$ and $4|t'|= |\Delta(0)-\Delta(\pi)|$. Here, $\Delta$ is the band gap, which can be negative after band inversion, at $k_z=0,\pi$ of a given valley. 

The crucial observation is that for $n=1$, $|t'|$ surpasses $|t|$ at $\Gamma$ for $m \ge 14$. When $n=2$, similar switch of the hopping strengths occurs for $m \ge 22$ . In contrast, for the superlattice series considered the thickness of SnTe layer is insufficient to cause switch of hopping strengths at the $M$ points. We then expect that the $(m,1)_{[111]}$ superlattice becomes a strong TI when $m\ge14$, and the $(m,2)_{[111]}$ superlattice becomes a $\mathcal{Z}_2$ strong TI when $m\ge22$. Indeed, this expectation is confirmed by computing $\mathcal{Z}_2$ invariant $\nu_0$ from the parities of Kohn-Sham Bloch wavefunctions \citep{SI}. The value of $\nu_0$ is 1 when the thickness of SnTe layer is beyond the thresholds, whereas below the threshold thickness of SnTe layer, we obtain $\nu_0=0$. For all superlattices with triple CaTe layer ($n=3,$ $m\le24$), our results show that they stay topologically trivial, consistent with the observation from Figs. \ref{fig:band111} (c) that there is no band inversion with the thickness of SnTe layers considered here.

\begin{figure}[ht]
\includegraphics[width=8 cm]{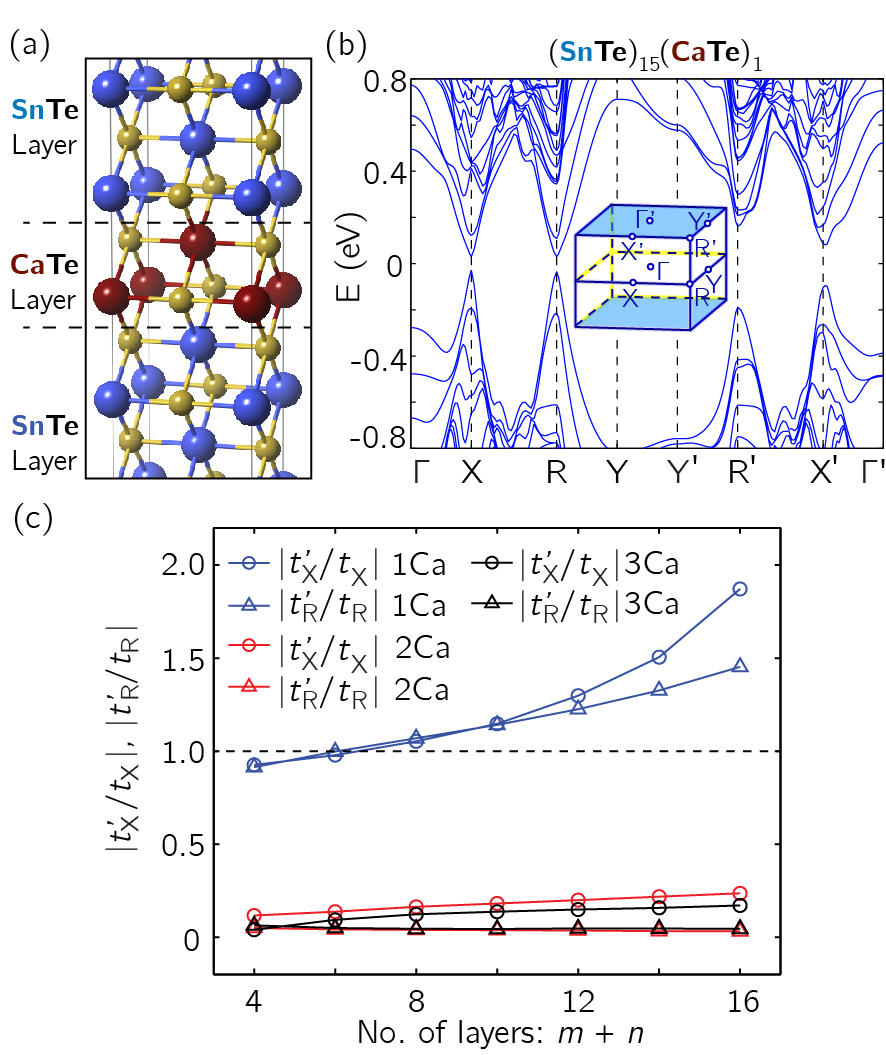}\caption{(a) A representative structure of $(m,2)_{[110]}$ superlattice. (b) A representative band structure near the band gap of $(15,1)_{[110]}$ superlattice. The inset shows the  Brillouin zone. (c) The values of $|t'/t|$  as a function of the thickness of superlattice, at $X$ and $R$. Blue, red and black curves correspond to $n=1$, 2 and 3.}
\label{fig:band110}
\end{figure}

We next consider a series of [110] superlattices with a single, double and triple layer of CaTe ($n=1,2,3$) (Fig. \ref{fig:band110}(a)). In the projected Brillouin zone along [110] direction,  there are  two symmetry-distinct Dirac valleys at  $\bar{X}(0, \pi)$ and $\bar{R}(\pi,\pi)$, respectively. Therefore, the [110] superlattices show direct band gaps at $X, X', R, R'$ (Fig.\ref{fig:band110} (b)). As shown in Fig. \ref{fig:band110} (c), for  $n=1$, band inversion occurs at both  $X$ and $R$ when $m \ge 7$.  Based on the calculation of the $\mathcal{Z}_2$ invariant, the  $(m,1)_{[110]}$ superlattices become a wTI. When $m+n=8,12$ and 16, the $\mathcal{Z}_2$ indices are $\nu_0;(\nu_1\nu_2\nu_3) = 0;(1, 0 ,0)$. When  $m+n=10$ and 14, the indices are 0;(1, 0, 1). In contrast, in the superlattices with  $n=2 $ and 3, the hopping through the spacer never exceeds that of the SnTe layers, and correspondingly, no phase transition is seen in the models calculated.

\textcolor{forestgreen}{\emph{\textsf{Discussion}.}}---
The SnTe/CaTe models have only uncovered a small region of the rich phase diagram, such as Figs.~\ref{fig:111}(b)-(c).
The identification of possible topological phases beyond the current system may be sought profitably in the variation of spacer material.
The spacer layer plays the critical role to mediate and modulate the valley-dependent couplings.
The tunability of coupling through the spacer can be elucidated based on a nearly-degenerate perturbation theory,
\begin{equation}
t'_v \approx - \frac{1}{2} \sum_k \left[ \frac{1}{E_{vk} - E_{v+}} + \frac{1}{E_{vk} - E_{v-}} \right] \check{t}_{k+}  \check{t}_{k-},
\end{equation}
where $E_{vk} - E_{v\tau_z}$ is the energy difference between the $k$th state of the spacer layer and the Dirac valley $(v,\tau_z)$, and $\check{t}_{k\tau_z}$ is the corresponding hopping amplitude. Anticipating the possibility of symmetry-breaking (e.g., ferroelectric transition \citep{Pawley66,Fu12}), we no longer require that the pair of Dirac interfacial states ($\tau_z=\pm$) be degenerate. As the energy differences enter as denominators, evidently the band gap of spacer layer and its alignment with the Dirac points become important parameters, the manipulation of which provides high tunability of $t'_v$. This can be achieved through materials choice. More interestingly, the level alignment can be changed through \emph{in situ} external gating, which may become an efficient experimental knob to tune the phase transitions in Figs.~\ref{fig:111}(b)-(c).

It is clear from our theoretical analysis and computational modeling that valley-dependent dimerization of Dirac interfacial states can be a powerful mechanism to design topological phases, out of superlattices of the same binary combination of materials. This mechanism is rather generic, and applies to any system with multiple band inversions and Dirac interfacial states, such as cTI, wTI and, quite possibly, elemental bismuth~\citep{Fu06, Fu07,Fu11, Fu12}. This strategy for tailoring valleys also have interesting implications in valleytronics~\citep{Rycerz07, Cao12, Xiao12, Li13}. One may envision valley valve or valley filter devices created by juxtaposition of heterostructures with massless and massive valleys tailored with the proposed mechanism. Elastic strain engineering will also find important applications here, where deformation can reversibly break the symmetry that relates a subset of valleys~\citep{Fang13}. Moreover, the ability to tune the number of Dirac surface states is especially attractive in the pursuit of novel Chern insulators~\citep{Fang13, Zhang13, Cai13}. It is also interesting to explore magnetic spacers in conjunction with band inversion to create superlattices with different Chern numbers.

\textcolor{forestgreen}{\emph{\textsf{Acknowledgements}.}}---
This work was supported by the National Science Foundation of China (NSFC Project 11174009) and China 973 Program Project 2013CB921900. FZ is supported by DARPA grant SPAWAR N66001-11-1-4110. We thank Profs. Junren Shi and Fa Wang for useful discussions.

\textcolor{forestgreen}{\emph{\textsf{Note added}.}}---{After this work was finalized, one complementary study by Yang et al~\citep{Yang13} demonstrated that wTI can emerge in a $[001]$ superlattice of SnTe/PbTe.}

\end{document}